\documentclass[aps,prl,twocolumn,showpacs,superscriptaddress,groupedaddress,floatfix]{revtex4-1}
\usepackage{latexsym}
\usepackage{graphicx}
\usepackage{dcolumn}
\usepackage{bm}
\usepackage{amssymb}
\usepackage{amsmath}
\usepackage[space]{grffile}
\usepackage{color}

\usepackage{subfig}

\begin{document}



\title{Central Charge of Periodically Driven Critical Kitaev Chains}
\date{\today}
\begin{abstract}
Periodically driven Kitaev chains show a rich phase diagram as the amplitude and frequency of the drive
is varied, with topological phase transitions separating regions with different number of Majorana zero and $\pi$ modes.
We explore whether the
critical point separating different phases of the periodically driven chain may be characterized by a universal central charge.
We affirmatively answer this question by studying the entanglement entropy (EE) numerically and analytically for the lowest entangled many particle 
eigenstate at arbitrary nonstroboscopic and stroboscopic times. 
We find that the EE at the critical point scales logarithmically with a time-independent central charge, and that the Floquet
micromotion gives only subleading corrections to the EE. This result also generalizes to multicritical points where the
EE is found to have a central charge that is the sum of the central charges of the intersecting critical lines.
\end{abstract}
\author{Daniel Yates}
\author{Yonah Lemonik}
\author{Aditi Mitra}

\affiliation{Center for Quantum Phenomena, Department of Physics, New York University, New York, NY, 10003, USA}
\maketitle


Periodic or Floquet driving has opened up new avenues of engineering correlated
quantum systems with behavior that is qualitatively different from
static systems~\cite{OkaRev18,Cayssol13}. As in equilibrium, we wish to have
universal descriptions of driven systems that do not depend on microscopic details. In equilibrium, critical states of matter 
possess a scale invariance that leads to such universal descriptions.  In one-dimensional (1D) static systems, this critical behavior can be
captured by conformal field theories (CFTs)~\cite{Vidal03,Calabrese04}. Do such universal descriptions exist for 1D Floquet systems?

To address this question, we study a 1D Floquet system, the periodically driven Kitaev chain with nearest neighbor (NN) and next-nearest neighbor (NNN) couplings~\cite{Kitaev01,Niu12,Sen13}. The static 
Kitaev chain has a ${Z}_2$ invariant, which is enlarged to a  ${Z}$ invariant with time reversal symmetry (TRS).
With driving, the system shows a rich phase diagram as the amplitude and frequency of the drive
is varied, with topological phase transitions separating regions with different numbers of Majorana modes~\cite{Yates17}.  
Moreover, the topological phases of the Floquet system is enhanced to
${Z}\times{Z}$ ~\cite{Kitagawa10,Rudner13, Delplace14, Asboth13,Roy17a}.

A universal characteristic of CFTs is their entanglement entropy (EE)~\cite{Calabrese04}. Further, entanglement spectra (ES)
(i.e, eigenvalues of the reduced density matrix) show an analogue of the  bulk-boundary correspondence of topological 
systems~\cite{Levin06,Preskill06}, and they are also sensitive to criticality~\cite{Casini09,Lemonik15}. 
In this paper, we explore the EE and ES of driven Floquet states. These quantities have the advantage that unlike 
thermodynamic quantities, the EE~\cite{Torre16} and ES extend naturally to nonequilibrium and driven systems, indeed to any quantum state.
However, there are several subtleties in thinking about the ES in the Floquet setting. The ES is a set of levels that 
span a range determined by the occupation probability of states, and thus, it has essentially the same appearance as the energy spectrum of a static Hamiltonian. However, 
in Floquet systems, energy is not conserved up to integer multiples of the drive frequency, so the conserved quasienergy is periodic. 
Thus, while there is one kind of zero mode in a static Hamiltonian and in the corresponding ES, 
there are \emph{two} kinds of such modes in a Floquet system: $0$ and $\pi$ modes. Since the ES is not periodic, there is no clear analog
of the $\pi$ mode in the ES~\cite{Yates16,Yates17}. 

A further wrinkle is that the ${Z}\times {Z}$ topological invariant and the quasienergy spectrum are properties of the full drive cycle, 
while the ES and EE are constructed from the instantaneous quantum state. They are therefore sensitive to which point in the drive cycle they are calculated.  
Thus, there is a conflict---one would expect that the ES and EE would carry information about the topological invariants; however, they are sensitive 
to within-cycle dynamics (also known as Floquet micromotion), which are not universal. 

Thus, it is unclear whether the
critical points separating different Floquet phases have any universal, time-independent description in terms of the EE, as static critical points do. 
In this paper, we find that the Floquet critical points \emph{do} have a universal form for the EE, despite the micromotion. 
In fact, they have precisely the same scaling law $S\sim \frac{c}{3}\log L$ as the static system, where $c$ is time independent, and depends on the number 
of $0$ \emph{and} $\pi$ modes. 
We also find equivalent behavior at multicritical points separating more than two phases~\cite{Berdanier18}.

We study the Kitaev chain with NN ($t_h,\Delta$) and NNN ($t_h',\Delta'$) tunneling and pairing
interactions. In terms of the complex fermion $c_i$ and its Fourier transformation $c_k$, the Hamiltonian is
\begin{align}
	\begin{split} \label{ham}
        H &= \sum_i \left[-t_h c_i^\dagger c_{i+1} -\Delta(t) c_i^\dagger c_{i+1}^\dagger
	-\mu(t) \left( c_{i}^\dagger c_i - \frac{1}{2}\right) \right. \\
	&\left. \qquad -t_h' c_i^\dagger c_{i+2} -\Delta'(t) c_i^\dagger c_{i+2}^\dagger + h.c. \right]
    \end{split}\nonumber \\
        &= \sum_k\begin{pmatrix} c_k^{\dagger}&c_{-k}\end{pmatrix}
H_{\rm BdG}(k,t) \begin{pmatrix} c_k\\c_{-k}^{\dagger}\end{pmatrix}.
\end{align}
The periodic driving may be applied to the chemical potential ($\mu$) or one or both of the pairing
amplitudes ($\Delta,\Delta'$). The results do not depend on
which parameter is varying in time.

In momentum space,
the Hamiltonian is $H_{\rm BdG}(k,t)= -\vec{d}(k,t)\cdot\vec{\sigma}$, where
$d_x(k,t)=0,d_y(k,t) = \Delta(t) \sin(k)+ \Delta'(t)\sin(2k),d_z(k,t)=t_h\cos(k)+t_h'\cos(2k)+\mu(t)/2 $.
For the numerical demonstrations, we drive both $\Delta$ and $\Delta'$, keeping $\mu$ static.
In units of $t_h = 1$, the parameters used are
$\Delta(t) = \Delta + 4 \sin (\Omega t)$, $\Delta'(t) = \Delta' + 4 \sin(\Omega t)$, $t_h' = -2, \Delta = 1, \Delta' = -2,
\Omega = 12$.

\begin{figure}
    \includegraphics[width = 0.5\textwidth, keepaspectratio]{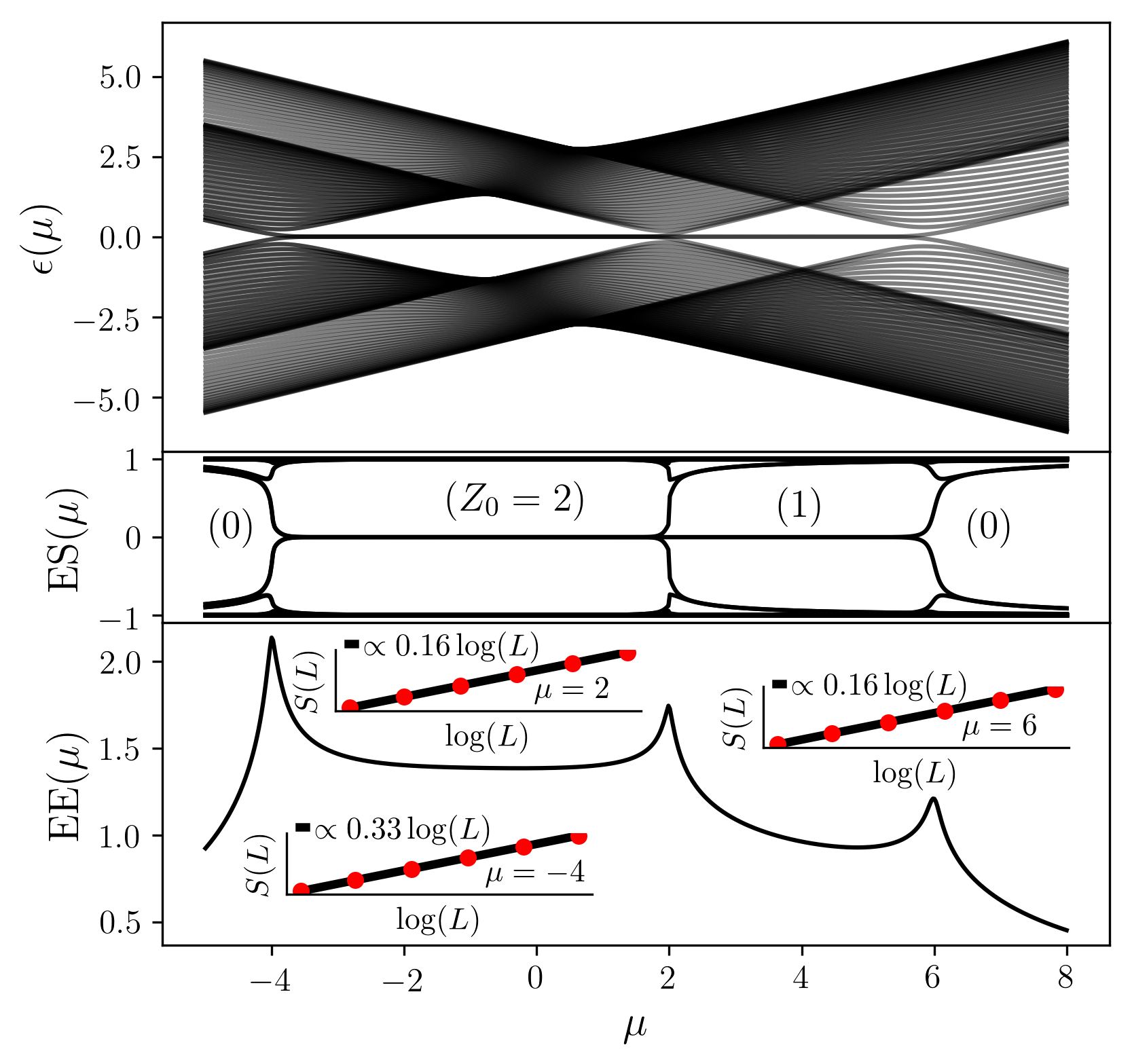}
    \caption{Static system, all plotted against $\mu$. Top panel, the energy levels of a wire with length $L = 75$.
Middle panel, the ES for an entanglement cut of length $L=75$ with periodic boundary conditions applied to the full density matrix.
    Bottom panel, the EE of the same. The insets in the bottom panel show how the EE (or S) at the critical points scale with 
$L$ with $400\leq L\leq 600$.} \label{static}
\end{figure}

The static Hamiltonian falls in the BDI classification~\cite{Ryu10}, with an integer
$Z$ characterizing the number of Majorana zero modes. This also equals the number of times 
the spinor  $\vec{d}(k)/|\vec{d}(k)|$ winds in the
$y$-$z$ plane in momentum space.
Figure~\ref{static} describes the static system. As $\mu$ is tuned, the system shows several topological phases.
These phases are distinguished by the number of Majorana zero modes in the energy spectrum (top
panel) and the ES (middle panel). In addition, the critical points separating the topological phases are characterized by an EE
that scales as (bottom panel)  $ S= (c/3)\log{L}$, where $L$ is the size of the subsystem associated with the reduced density matrix,
and $c$ is the central charge. For a critical point
separating a phase with $Z$ Majorana modes from one with $Z'$ Majorana modes,
the numerically extracted central charge is $c = |Z-Z'|/2$~\cite{Pollmann17}.
In this paper, we wish to understand how this fundamental result for the scaling of the EE of critical static phases
generalizes to critical Floquet phases.

\begin{figure}
    \includegraphics[width = 0.5\textwidth, keepaspectratio]{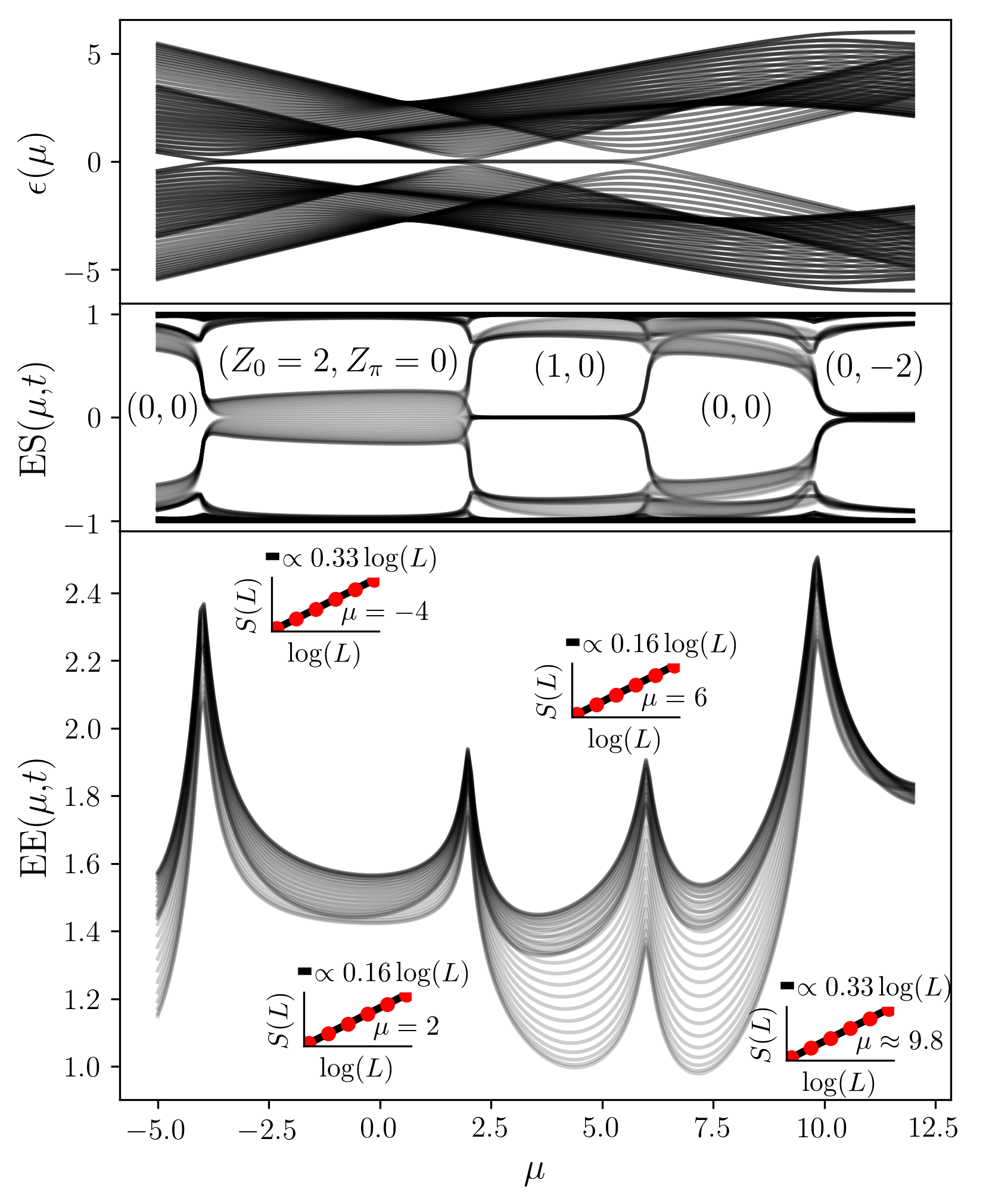}
    \caption{Floquet system, all plotted against $\mu$. Top panel, the quasienergy levels of the wire
    with size $L = 50$. The $\pi$ modes are visible at the FBZ boundary $|\epsilon| = \Omega/2 = 6$ for $\mu>10$. 
    Middle panel, the ES at several different times within a period (different solid lines) 
    for an entanglement cut of size $L=50$. The strongest time dependences are at
    zero entanglement energies. Bottom panel, the corresponding 
    time dependent EE (solid lines are for different times
    within a period). The insets in the bottom
    panel show how the EE(or S) at the critical points scale with $L$ with
    $400\leq L\leq 600$.
    The leading logarithmic contribution at the critical points is time invariant. 
} \label{fgs}
\end{figure}

\begin{figure}
\includegraphics[width=0.5\textwidth,keepaspectratio]{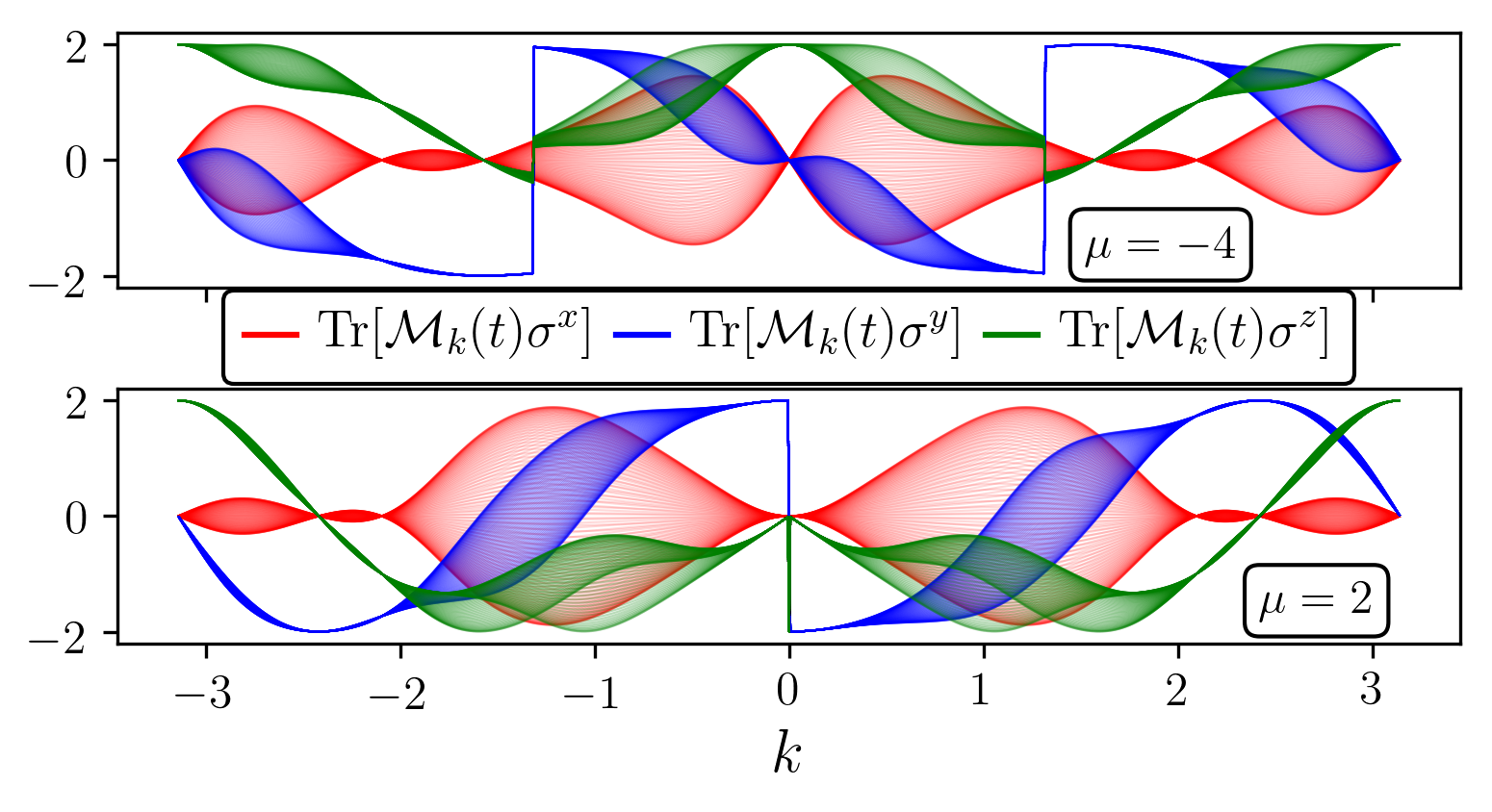}
    \caption{Discontinuities in $\mathcal{M}_k(t)$ for the FGS,
    for several times during a driving period (different solid lines) and at two different critical $\mu$.
    The discontinuities send the Bloch-vector to the opposite side of the sphere at
    all times, with the orientation of the jump varying in time. Number of discontinuous eigenvalues of ${\cal M}_k$ 
are $N_T=4$ (top) and $N_T=2$ (bottom).
The discontinuities in the $\sigma_x$ projection are difficult to see. 
Away from the critical $\mu$ values (not shown),
    all of the projections are continuous.}\label{fgsjumps}
\end{figure}

\begin{figure}
    \includegraphics[width=0.5\textwidth,keepaspectratio]{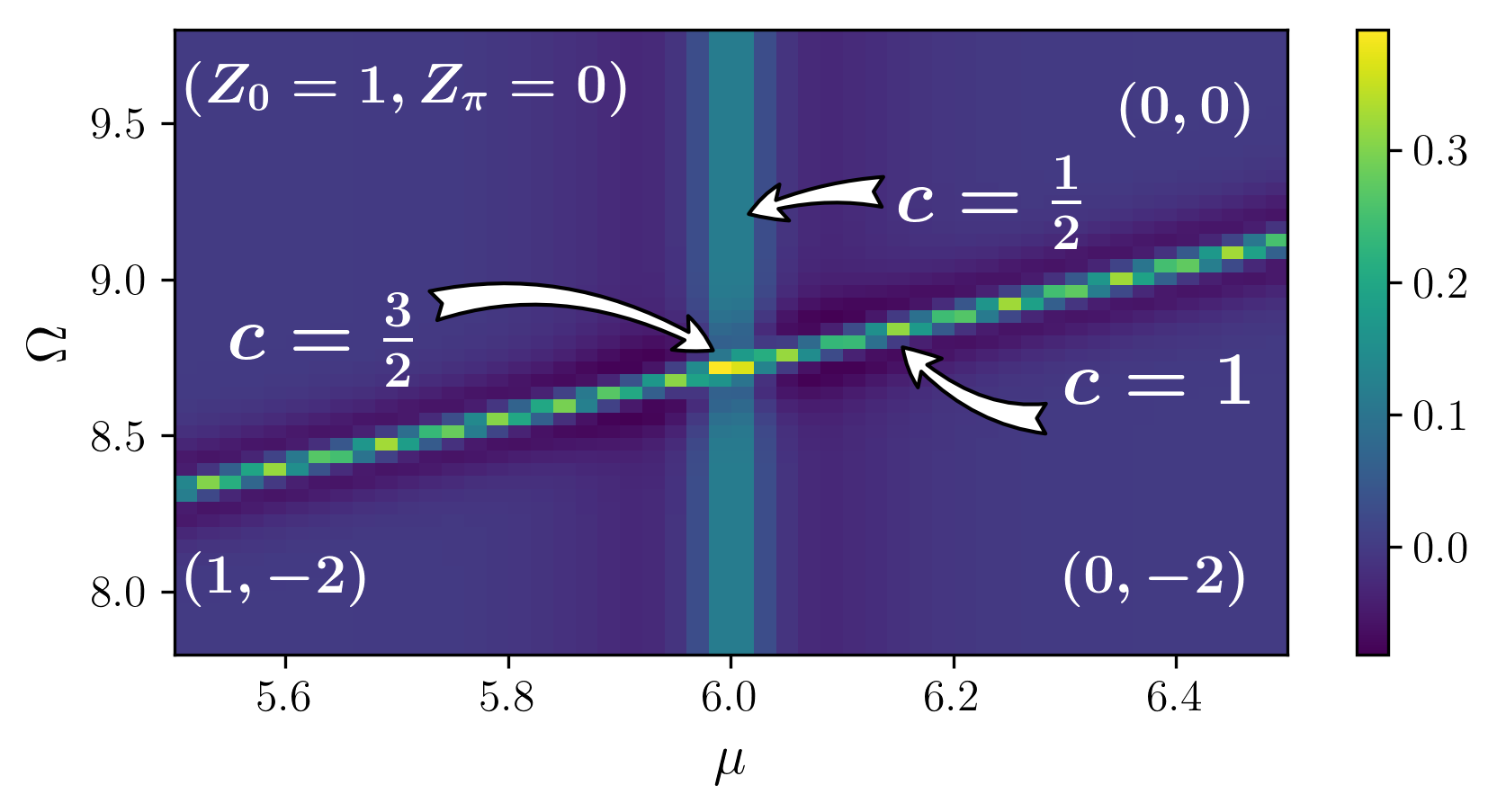}
    \caption{Phase diagram of the prefactor to $\log (L)$ in the EE scaling for the FGS,
    as a function of $\mu$ and $\Omega$. The multicritical point separates the four phases $(Z_0,Z_{\pi})=(0,0),(1,0),(1,-2),(0,-2)$.
    The leading logarithmic scaling at the critical lines and multicritical point are time independent.}\label{multi}
\end{figure}

In particular, we are interested in the entanglement scaling of the Floquet
ground state (FGS),
which is a half filled many-body eigenstate of
the Floquet Hamiltonian $H_F= H(t)-i\partial_t$. This eigenstate is a Slater determinant of the time periodic Floquet modes  $|\phi(k,t)\rangle$,
defined as the eigenmodes of $H_F$, $H_F|\phi(k,t)\rangle =\epsilon_k |\phi(k,t)\rangle$. $\epsilon_k$ are
the quasienergies, and they are restricted within a Floquet Brillouin zone (FBZ) of size $\Omega$~\cite{Shirley65,Sambe73}.
The half filled state corresponding to the FGS is such as to ensure area law scaling of the EE when the system has a
gap in the quasienergy spectrum. Concretely, restricting the quasienergy spectrum to lie between $-\Omega/2,\Omega/2$, and noting that
the chiral symmetry of the Floquet Hamiltonian causes the quasienergy spectra to come in pairs of
$\pm|\epsilon_k|$, the FGS  corresponds to occupying with probability $1$ all Floquet modes with negative quasienergy.
This should be contrasted with a half filled state obtained from unitary time evolution under $H(t)$ from an arbitrary initial state,
where such a state will show volume law scaling of the EE at a steady state~\cite{Yates16,Yates17}.

We briefly explain how the ES and EE are studied numerically and analytically. The underlying principle is that for a system of free fermions,
the eigenvalues of the reduced density matrix can be extracted from the eigenvalues of only the two-point correlation function, a consequence
of Wick's theorem~\cite{Eisler2009,RMPVedral}. The relevant correlation matrix for our half filled state is
\begin{align}\label{corr}
G_{i,j}(t)=\int_{-\pi}^{\pi}\frac{dk}{2\pi}e^{i k (i-j)}\mathcal{M}_k(t),
\end{align}
where $i,j$ index the physical sites within the entanglement cut, $\mathcal{M}$ is a $2\times 2$ matrix, which for the
static ground state and FGS, are respectively,
\begin{align}
\mathcal{M}_{k,\rm static} =
    \frac{\vec{d}(k)\cdot\vec{\sigma}}{|d(k)|}; \mathcal{M}_{k,\rm FGS}(t) =\langle \phi(k,t)|\vec{\sigma}|\phi(k,t)\rangle\cdot\vec{\sigma}.
\end{align}
$G_{ij}$ is a Hermitian matrix whose expansion, in terms of 
Pauli  matrices, implies that eigenvalues come in pairs $\pm \lambda_i$, giving an EE,
\begin{align}
&S=-\frac{1}{2}\sum_{\alpha=\pm,\lambda_i}\biggl[\left(\frac{1-\alpha \lambda_i}{2}\right)\ln\left(\frac{1-\alpha\lambda_i}{2}\right)
\biggr].
\end{align}
The Majorana modes in the ES are pinned exactly at zero entanglement energies (middle panel, Fig.~\ref{static}).

There are some key differences between static and Floquet topological phases.
In the presence of Floquet driving, the definition of TRS is subtle. There are two TRS points $t^*$ within a cycle where the
Hamiltonian obeys $H(t+t^*)= H(-t+t^*)$ for all $t$. For our drive, these are $t^*=\pi/2\Omega,3\pi/2\Omega$.

The quasienergy spectrum
hosts Majorana modes that are either pinned at zero quasienergy, or at the Floquet zone boundaries. We will denote the former as
Majorana zero modes (MZM) and the latter by Majorana $\pi$ modes (MPM). The Floquet phase is now
characterized by $Z_0\times Z_{\pi}$, where $Z_0(Z_{\pi})$ refers to the  number of MZMs (MPMs).
Figure~\ref{fgs} (top panel) displays the quasienergy levels for the time periodic chain. As $\mu$ is increased, several transitions are visible, 
going from trivial to 2MZM to 1MZM to trivial to 2MPM.

Since quasienergies are not sensitive to the micromotion, while EE and ES are, this leads to some ambiguity between the topological
characterization via the quasienergy, and that from the entanglement.
The topological phase transitions are visible in the ES (middle panel) in a different way. First, the
ES is characterized by a single gap, and all edge modes have to lie within this gap.
In addition,
the winding of the Floquet states in momentum space is, strictly speaking, well-defined only at the two TRS times.
At other times
of the drive, the Floquet modes acquire a nonzero projection along all three directions $\hat{x},\hat{y},\hat{z}$ so that the winding is ill defined. 
This leads to an ES where the Majorana zero (entanglement) energy modes appear only at the two
discrete times $t^*$ in the ES, while at other times, the Majorana modes on the same sides of the entanglement cut couple to each other, forming
complex fermions. Although these complex fermions are still localized at the entanglement cuts, their entanglement energies are no
longer pinned at zero.
Thus, while at the two TRS times, the number of Majorana modes in the ES are $|Z_0\pm Z_{\pi}|$ respectively~\cite{Suppmat}, at other times,
the ES shows a $Z_2$ invariance. The reason for $Z_2$ is that
if there are an odd number of Majorana modes at an entanglement cut, one unpaired Majorana mode persists when $t\neq t^*$.
This physics is highlighted in Fig.~\ref{fgs} (middle-panel), where the ES, through a series of topological phases obtained from varying $\mu$,
is shown at several different times of the drive cycle. The time-dependence is the strongest at zero entanglement energies~\cite{Suppmat}, with
the zero modes appearing only at special times $t^*$.

What is remarkable is that the EE (bottom panel Fig.~\ref{fgs}) constructed out of this ES, despite
the fact that the zero modes exist at only two discrete times during a cycle,
still scales logarithmically at the critical points with a time independent central charge.
Note that, at all points, including the critical points, the EE is time dependent. This makes the time-independent central charge nontrivial.
The time dependence from micromotion only gives subleading corrections, in size of the entanglement cut $L$, to the EE at the critical point.
In contrast, away from the critical points, due to the presence of the gap, the area law holds. In this case, the micromotion affects the EE
to leading order. This is apparent in the bottom panel of Fig.~\ref{fgs}, where the time dependence of the EE is largest away from the critical points.

Reference~\cite{Suppmat} shows how the EE scales as one crosses the several topological phases
as a function of time and system size.
Irrespective of the micromotion, the entanglement scales  as in a static critical phase but with a modified central charge,
\begin{eqnarray}
c = \biggl(|Z_0- Z_0'|+|Z_{\pi}-Z_{\pi}'|\biggr)/2~\label{cfl},
\end{eqnarray}
with any deviations from the above decreasing with momentum space resolution.

We explain this robust central charge as follows.
The logarithmic scaling originates from a discontinuity in the matrix $\mathcal{M}_k$.
For example, for noninteracting complex
fermions ($\Delta=\Delta'=0$), $\mathcal{M}_k$ is a scalar with a step function at the Fermi momentum. This leads to a power law $G_{ij}\sim 1/|i-j|$
in the correlation function and an EE that scales with $(1/3)\ln{L}$, and hence, $c=1$~\cite{Wolf06,Klich06}.
For the BdG Hamiltonians under consideration here,
the discontinuity is reflected in special $k$ points, where the dispersion $\epsilon(k^*)=0$ and $\mathcal{M}({k^{*+}})\neq \mathcal{M}({k^{*-}})$\cite{ares15}. For example,
for $\Delta'=0,t_h'=0,t_h=\Delta,\mu=2t_h$, $\mathcal{M}_k=\frac{\cos(k/2)}{|\cos(k/2)|}\left[\cos(k/2)\sigma_z + \sin(k/2)\sigma_y\right]$.
The dispersion vanishes at $k^*=\pi$, and around this point, $\mathcal{M}_k$ has the discontinuity
$\mathcal{M}(\pi^+)=\sigma_y, \mathcal{M}(\pi^-)=-\sigma_y$.
This discontinuity gives rise to power-law correlations in position, and a corresponding EE that scales as 
$S=(c/3)\ln{L}$ with $c=1/2$~\cite{Suppmat}.

Consider another example with NNN terms that can give rise to multiple Majorana modes.
For $\Delta=\Delta'= t_h=t_h', \mu= 2 t_h$,
$\mathcal{M}_k=\frac{1+ 2\cos(k)}{|1+2 \cos(k)|}\left[\cos(k)\sigma_z + \sin(k)\sigma_y\right] $. The dispersion now vanishes at
two points in momentum space
corresponding to $k^*=\pm 2\pi/3$. Across these $k^*$, the $\mathcal{M}_k$ are discontinuous as follows,
$\mathcal{M}(k^{*+})= \sigma_y =-\mathcal{M}(k^{*-})$. Each of these points gives a central charge of $1/2$, implying
a total central charge of $c=1$. Thus, quite simply, the total central charge is $c=N_T/4$, where $N_T$ is the number of
discontinuous eigenvalues of $\mathcal{M}_k$.
Reference~\cite{Suppmat} demonstrates these discontinuities at the critical points of the static system shown in Fig.~\ref{static}. 

Similar to the static case, the central charge of the Floquet system follows from the nature of the discontinuities in the 
$\mathcal{M}_{k,\rm FGS}$.
Figure~\ref{fgsjumps} (and Ref.~\cite{Suppmat}) shows that despite the micromotion of the Floquet states,
$\mathcal{M}_{k,\rm FGS}$ maintains a time-independent jump across momenta $k^*$ at which the quasienergy vanishes. This fact holds for
both changes in $Z_0$ and/or $Z_{\pi}$ at the transition. 
The origin of the discontinuity is that the FGS is
constructed from ``filling'' all quasienergy levels of the same band, introducing a ``Fermi'' point in momentum space.
This discontinuity can again be indexed by the number of discontinuous eigenvalues $N_T$.
Figure~\ref{fgsjumps} plots $\mathcal{M}_{k,\rm FGS}$ projected onto the Pauli matrices for many times during the drive cycle and for several
different Floquet critical points. We find that $N_T=2\left(|Z_0- Z_0'|+|Z_{\pi}-Z_{\pi}'|\right)$.
The time dependence only changes the location of the jump on the Bloch sphere.
While clearly the leading scaling of the EE is like that of a static critical
theory with a well-defined central-charge, the EE does show periodicity in time.
This periodic behavior only affects the subleading behavior in the EE at the critical point.

We now give analytic arguments for the numerical results.
Expanding around $k=k^*$ where the dispersion vanishes, and therefore $\mathcal{M}_k$ is singular, we write,
\begin{eqnarray}
\mathcal{M}_{k,\rm FGS}\simeq \frac{(k-k^*)}{|k-k^*|}{\sigma}_{1}(t) + \vec{g}_k(t) \cdot \vec{\sigma}\label{Mfgs},
\end{eqnarray}
where ${\sigma}_{1}(t)=\hat{n}(t)\cdot\vec{\sigma}$ with $\hat{n}$ a unit vector. The discontinuous prefactor
contains the physics of the ``Fermi'' point associated with the FGS. In contrast, $\vec{g}_k$ is a
smooth function of $k$. The time dependence of $\vec{g}_k,\sigma_{1}$ are due to Floquet micromotion.
In the static problems~\cite{Peschel04}, $\sigma_{1}=\sigma_y$. Regardless of the value of ${\sigma}_1(t)$,
as one crosses $k^*$, the matrix jumps from $\sigma_1$ to $-{\sigma}_1$, and $N_T=2$ at all times. Equation~\eqref{Mfgs} is
valid whether we have jumps in $Z_0$ and/or $Z_{\pi}$, where the difference between the two kinds of modes
is encoded in the micromotion, i.e., the precise time dependence of $\sigma_1(t),g(t)$. 
 
The Fourier transformation of Eq.~\eqref{Mfgs} is
\begin{eqnarray}
&&G_{ij}(t)\sim i\frac{e^{ik^*(i-j)}}{\pi (i-j)}{\sigma}_{1}(t) \nonumber\\
&&+ \vec{g}_{k=0}(t)\cdot\vec{\sigma} \delta(i-j)+\vec{g}'_{k=0}(t)\cdot\vec{\sigma}\delta'(i-j)
+ \ldots.\label{Gijf}
\end{eqnarray}
Thus, the smooth function $\vec{g}_k$ gives only short-ranged correlations.
The discontinuity at $k=k^*$, despite the oscillation
$e^{ik^*(i-j)}$, gives~\cite{Suppmat} logarithmic scaling
of the EE. When there are many ``Fermi'' points, the EE from each
singular point combines additively. Note that, one cannot rule out nontopological
gap closings, in which case Eq.~\eqref{cfl} provides a lower bound.

Floquet micromotion only affects short distance correlations because the
micromotion is over a time $t\leq \Omega^{-1}$, and it is therefore associated with
a finite spatial range $t_h/\Omega$ in units of the lattice spacing. The short distance physics 
cannot affect the power-law tail of Eq.~\eqref{Gijf}, which extends over arbitrary long distances.
However, when the system is gapped,
and the correlations are short ranged, then the micromotion is the leading correction, giving a strong time dependence to
the EE (Fig.~\ref{fgs}). 

The richness of phases under a periodic drive leads not only to critical points separating two different phases but also multicritical
points. Figure~\ref{multi} shows a multicritical point separating four phases. This multicritical point is the meeting point of two critical lines,
and it is associated with a central charge $c=c_1+c_2$,
where $c_{1,2}$ are the central charges of the two intersecting critical lines. For the example shown, $c=3/2=1+1/2$.

We have shown that a critical (multicritical) point separating two (or more)
Floquet phases, despite the time dependence,
has a universal behavior for the EE; namely that it scales as $(c/3)\ln{L}$, where the central charge accounts for 
MZMs and MPMs [Eq.~\eqref{cfl}].
The time dependence due to micromotion gives subleading corrections that obey the area law. Away from the critical point, these
subleading corrections become the dominant correction, and the EE shows a strong time dependence. 
How these results are affected by interactions is an interesting open question. 

{\sl Acknowledgements:}
This work was supported by the US Department of Energy, Office of Science,
Basic Energy Sciences, under Award No.~DE-SC0010821.


\begin{thebibliography}{30}%
\makeatletter
\providecommand \@ifxundefined [1]{%
 \@ifx{#1\undefined}
}%
\providecommand \@ifnum [1]{%
 \ifnum #1\expandafter \@firstoftwo
 \else \expandafter \@secondoftwo
 \fi
}%
\providecommand \@ifx [1]{%
 \ifx #1\expandafter \@firstoftwo
 \else \expandafter \@secondoftwo
 \fi
}%
\providecommand \natexlab [1]{#1}%
\providecommand \enquote  [1]{``#1''}%
\providecommand \bibnamefont  [1]{#1}%
\providecommand \bibfnamefont [1]{#1}%
\providecommand \citenamefont [1]{#1}%
\providecommand \href@noop [0]{\@secondoftwo}%
\providecommand \href [0]{\begingroup \@sanitize@url \@href}%
\providecommand \@href[1]{\@@startlink{#1}\@@href}%
\providecommand \@@href[1]{\endgroup#1\@@endlink}%
\providecommand \@sanitize@url [0]{\catcode `\\12\catcode `\$12\catcode
  `\&12\catcode `\#12\catcode `\^12\catcode `\_12\catcode `\%12\relax}%
\providecommand \@@startlink[1]{}%
\providecommand \@@endlink[0]{}%
\providecommand \url  [0]{\begingroup\@sanitize@url \@url }%
\providecommand \@url [1]{\endgroup\@href {#1}{\urlprefix }}%
\providecommand \urlprefix  [0]{URL }%
\providecommand \Eprint [0]{\href }%
\providecommand \doibase [0]{http://dx.doi.org/}%
\providecommand \selectlanguage [0]{\@gobble}%
\providecommand \bibinfo  [0]{\@secondoftwo}%
\providecommand \bibfield  [0]{\@secondoftwo}%
\providecommand \translation [1]{[#1]}%
\providecommand \BibitemOpen [0]{}%
\providecommand \bibitemStop [0]{}%
\providecommand \bibitemNoStop [0]{.\EOS\space}%
\providecommand \EOS [0]{\spacefactor3000\relax}%
\providecommand \BibitemShut  [1]{\csname bibitem#1\endcsname}%
\let\auto@bib@innerbib\@empty
\bibitem [{\citenamefont {Oka}\ and\ \citenamefont
  {Kitamura}(shed)}]{OkaRev18}%
  \BibitemOpen
  \bibfield  {author} {\bibinfo {author} {\bibfnamefont {T.}~\bibnamefont
  {Oka}}\ and\ \bibinfo {author} {\bibfnamefont {S.}~\bibnamefont {Kitamura}},\
  }\href@noop {} {\bibfield  {journal} {\bibinfo  {journal} {arXiv:1804.03212}\
  } (\bibinfo {year} {unpublished})}\BibitemShut {NoStop}%
\bibitem [{\citenamefont {Cayssol}\ \emph {et~al.}(2013)\citenamefont
  {Cayssol}, \citenamefont {D\'ora}, \citenamefont {Simon},\ and\ \citenamefont
  {Moessner}}]{Cayssol13}%
  \BibitemOpen
  \bibfield  {author} {\bibinfo {author} {\bibfnamefont {J.}~\bibnamefont
  {Cayssol}}, \bibinfo {author} {\bibfnamefont {B.}~\bibnamefont {D\'ora}},
  \bibinfo {author} {\bibfnamefont {F.}~\bibnamefont {Simon}}, \ and\ \bibinfo
  {author} {\bibfnamefont {R.}~\bibnamefont {Moessner}},\ }\href@noop {}
  {\bibfield  {journal} {\bibinfo  {journal} {Phys. Status Solidi RRL}\
  }\textbf {\bibinfo {volume} {7}},\ \bibinfo {pages} {101} (\bibinfo {year}
  {2013})}\BibitemShut {NoStop}%
\bibitem [{\citenamefont {Vidal}\ \emph {et~al.}(2003)\citenamefont {Vidal},
  \citenamefont {Latorre}, \citenamefont {Rico},\ and\ \citenamefont
  {Kitaev}}]{Vidal03}%
  \BibitemOpen
  \bibfield  {author} {\bibinfo {author} {\bibfnamefont {G.}~\bibnamefont
  {Vidal}}, \bibinfo {author} {\bibfnamefont {J.~I.}\ \bibnamefont {Latorre}},
  \bibinfo {author} {\bibfnamefont {E.}~\bibnamefont {Rico}}, \ and\ \bibinfo
  {author} {\bibfnamefont {A.}~\bibnamefont {Kitaev}},\ }\href {\doibase
  10.1103/PhysRevLett.90.227902} {\bibfield  {journal} {\bibinfo  {journal}
  {Phys. Rev. Lett.}\ }\textbf {\bibinfo {volume} {90}},\ \bibinfo {pages}
  {227902} (\bibinfo {year} {2003})}\BibitemShut {NoStop}%
\bibitem [{\citenamefont {Calabrese}\ and\ \citenamefont
  {Cardy}(2004)}]{Calabrese04}%
  \BibitemOpen
  \bibfield  {author} {\bibinfo {author} {\bibfnamefont {P.}~\bibnamefont
  {Calabrese}}\ and\ \bibinfo {author} {\bibfnamefont {J.}~\bibnamefont
  {Cardy}},\ }\href {http://stacks.iop.org/1742-5468/2004/i=06/a=P06002}
  {\bibfield  {journal} {\bibinfo  {journal} {Journal of Statistical Mechanics:
  Theory and Experiment}\ }\textbf {\bibinfo {volume} {2004}},\ \bibinfo
  {pages} {P06002} (\bibinfo {year} {2004})}\BibitemShut {NoStop}%
\bibitem [{\citenamefont {Kitaev}(2001)}]{Kitaev01}%
  \BibitemOpen
  \bibfield  {author} {\bibinfo {author} {\bibfnamefont {A.~Y.}\ \bibnamefont
  {Kitaev}},\ }\href {http://stacks.iop.org/1063-7869/44/i=10S/a=S29}
  {\bibfield  {journal} {\bibinfo  {journal} {Physics-Uspekhi}\ }\textbf
  {\bibinfo {volume} {44}},\ \bibinfo {pages} {131} (\bibinfo {year}
  {2001})}\BibitemShut {NoStop}%
\bibitem [{\citenamefont {Niu}\ \emph {et~al.}(2012)\citenamefont {Niu},
  \citenamefont {Chung}, \citenamefont {Hsu}, \citenamefont {Mandal},
  \citenamefont {Raghu},\ and\ \citenamefont {Chakravarty}}]{Niu12}%
  \BibitemOpen
  \bibfield  {author} {\bibinfo {author} {\bibfnamefont {Y.}~\bibnamefont
  {Niu}}, \bibinfo {author} {\bibfnamefont {S.~B.}\ \bibnamefont {Chung}},
  \bibinfo {author} {\bibfnamefont {C.-H.}\ \bibnamefont {Hsu}}, \bibinfo
  {author} {\bibfnamefont {I.}~\bibnamefont {Mandal}}, \bibinfo {author}
  {\bibfnamefont {S.}~\bibnamefont {Raghu}}, \ and\ \bibinfo {author}
  {\bibfnamefont {S.}~\bibnamefont {Chakravarty}},\ }\href {\doibase
  10.1103/PhysRevB.85.035110} {\bibfield  {journal} {\bibinfo  {journal} {Phys.
  Rev. B}\ }\textbf {\bibinfo {volume} {85}},\ \bibinfo {pages} {035110}
  (\bibinfo {year} {2012})}\BibitemShut {NoStop}%
\bibitem [{\citenamefont {Thakurathi}\ \emph {et~al.}(2013)\citenamefont
  {Thakurathi}, \citenamefont {Patel}, \citenamefont {Sen},\ and\ \citenamefont
  {Dutta}}]{Sen13}%
  \BibitemOpen
  \bibfield  {author} {\bibinfo {author} {\bibfnamefont {M.}~\bibnamefont
  {Thakurathi}}, \bibinfo {author} {\bibfnamefont {A.~A.}\ \bibnamefont
  {Patel}}, \bibinfo {author} {\bibfnamefont {D.}~\bibnamefont {Sen}}, \ and\
  \bibinfo {author} {\bibfnamefont {A.}~\bibnamefont {Dutta}},\ }\href
  {\doibase 10.1103/PhysRevB.88.155133} {\bibfield  {journal} {\bibinfo
  {journal} {Phys. Rev. B}\ }\textbf {\bibinfo {volume} {88}},\ \bibinfo
  {pages} {155133} (\bibinfo {year} {2013})}\BibitemShut {NoStop}%
\bibitem [{\citenamefont {Yates}\ and\ \citenamefont {Mitra}(2017)}]{Yates17}%
  \BibitemOpen
  \bibfield  {author} {\bibinfo {author} {\bibfnamefont {D.~J.}\ \bibnamefont
  {Yates}}\ and\ \bibinfo {author} {\bibfnamefont {A.}~\bibnamefont {Mitra}},\
  }\href {\doibase 10.1103/PhysRevB.96.115108} {\bibfield  {journal} {\bibinfo
  {journal} {Phys. Rev. B}\ }\textbf {\bibinfo {volume} {96}},\ \bibinfo
  {pages} {115108} (\bibinfo {year} {2017})}\BibitemShut {NoStop}%
\bibitem [{\citenamefont {Kitagawa}\ \emph {et~al.}(2010)\citenamefont
  {Kitagawa}, \citenamefont {Berg}, \citenamefont {Rudner},\ and\ \citenamefont
  {Demler}}]{Kitagawa10}%
  \BibitemOpen
  \bibfield  {author} {\bibinfo {author} {\bibfnamefont {T.}~\bibnamefont
  {Kitagawa}}, \bibinfo {author} {\bibfnamefont {E.}~\bibnamefont {Berg}},
  \bibinfo {author} {\bibfnamefont {M.}~\bibnamefont {Rudner}}, \ and\ \bibinfo
  {author} {\bibfnamefont {E.}~\bibnamefont {Demler}},\ }\href {\doibase
  10.1103/PhysRevB.82.235114} {\bibfield  {journal} {\bibinfo  {journal} {Phys.
  Rev. B}\ }\textbf {\bibinfo {volume} {82}},\ \bibinfo {pages} {235114}
  (\bibinfo {year} {2010})}\BibitemShut {NoStop}%
\bibitem [{\citenamefont {Rudner}\ \emph {et~al.}(2013)\citenamefont {Rudner},
  \citenamefont {Lindner}, \citenamefont {Berg},\ and\ \citenamefont
  {Levin}}]{Rudner13}%
  \BibitemOpen
  \bibfield  {author} {\bibinfo {author} {\bibfnamefont {M.~S.}\ \bibnamefont
  {Rudner}}, \bibinfo {author} {\bibfnamefont {N.~H.}\ \bibnamefont {Lindner}},
  \bibinfo {author} {\bibfnamefont {E.}~\bibnamefont {Berg}}, \ and\ \bibinfo
  {author} {\bibfnamefont {M.}~\bibnamefont {Levin}},\ }\href {\doibase
  10.1103/PhysRevX.3.031005} {\bibfield  {journal} {\bibinfo  {journal} {Phys.
  Rev. X}\ }\textbf {\bibinfo {volume} {3}},\ \bibinfo {pages} {031005}
  (\bibinfo {year} {2013})}\BibitemShut {NoStop}%
\bibitem [{\citenamefont {Asb\'oth}\ \emph {et~al.}(2014)\citenamefont
  {Asb\'oth}, \citenamefont {Tarasinski},\ and\ \citenamefont
  {Delplace}}]{Delplace14}%
  \BibitemOpen
  \bibfield  {author} {\bibinfo {author} {\bibfnamefont {J.~K.}\ \bibnamefont
  {Asb\'oth}}, \bibinfo {author} {\bibfnamefont {B.}~\bibnamefont
  {Tarasinski}}, \ and\ \bibinfo {author} {\bibfnamefont {P.}~\bibnamefont
  {Delplace}},\ }\href {\doibase 10.1103/PhysRevB.90.125143} {\bibfield
  {journal} {\bibinfo  {journal} {Phys. Rev. B}\ }\textbf {\bibinfo {volume}
  {90}},\ \bibinfo {pages} {125143} (\bibinfo {year} {2014})}\BibitemShut
  {NoStop}%
\bibitem [{\citenamefont {Asb\'oth}\ and\ \citenamefont
  {Obuse}(2013)}]{Asboth13}%
  \BibitemOpen
  \bibfield  {author} {\bibinfo {author} {\bibfnamefont {J.~K.}\ \bibnamefont
  {Asb\'oth}}\ and\ \bibinfo {author} {\bibfnamefont {H.}~\bibnamefont
  {Obuse}},\ }\href {\doibase 10.1103/PhysRevB.88.121406} {\bibfield  {journal}
  {\bibinfo  {journal} {Phys. Rev. B}\ }\textbf {\bibinfo {volume} {88}},\
  \bibinfo {pages} {121406} (\bibinfo {year} {2013})}\BibitemShut {NoStop}%
\bibitem [{\citenamefont {Roy}\ and\ \citenamefont {Harper}(2017)}]{Roy17a}%
  \BibitemOpen
  \bibfield  {author} {\bibinfo {author} {\bibfnamefont {R.}~\bibnamefont
  {Roy}}\ and\ \bibinfo {author} {\bibfnamefont {F.}~\bibnamefont {Harper}},\
  }\href {\doibase 10.1103/PhysRevB.96.155118} {\bibfield  {journal} {\bibinfo
  {journal} {Phys. Rev. B}\ }\textbf {\bibinfo {volume} {96}},\ \bibinfo
  {pages} {155118} (\bibinfo {year} {2017})}\BibitemShut {NoStop}%
\bibitem [{\citenamefont {Levin}\ and\ \citenamefont {Wen}(2006)}]{Levin06}%
  \BibitemOpen
  \bibfield  {author} {\bibinfo {author} {\bibfnamefont {M.}~\bibnamefont
  {Levin}}\ and\ \bibinfo {author} {\bibfnamefont {X.-G.}\ \bibnamefont
  {Wen}},\ }\href {\doibase 10.1103/PhysRevLett.96.110405} {\bibfield
  {journal} {\bibinfo  {journal} {Phys. Rev. Lett.}\ }\textbf {\bibinfo
  {volume} {96}},\ \bibinfo {pages} {110405} (\bibinfo {year}
  {2006})}\BibitemShut {NoStop}%
\bibitem [{\citenamefont {Kitaev}\ and\ \citenamefont
  {Preskill}(2006)}]{Preskill06}%
  \BibitemOpen
  \bibfield  {author} {\bibinfo {author} {\bibfnamefont {A.}~\bibnamefont
  {Kitaev}}\ and\ \bibinfo {author} {\bibfnamefont {J.}~\bibnamefont
  {Preskill}},\ }\href {\doibase 10.1103/PhysRevLett.96.110404} {\bibfield
  {journal} {\bibinfo  {journal} {Phys. Rev. Lett.}\ }\textbf {\bibinfo
  {volume} {96}},\ \bibinfo {pages} {110404} (\bibinfo {year}
  {2006})}\BibitemShut {NoStop}%
\bibitem [{\citenamefont {Casini}\ and\ \citenamefont
  {Huerta}(2009)}]{Casini09}%
  \BibitemOpen
  \bibfield  {author} {\bibinfo {author} {\bibfnamefont {H.}~\bibnamefont
  {Casini}}\ and\ \bibinfo {author} {\bibfnamefont {M.}~\bibnamefont
  {Huerta}},\ }\href {http://stacks.iop.org/1751-8121/42/i=50/a=504007}
  {\bibfield  {journal} {\bibinfo  {journal} {Journal of Physics A:
  Mathematical and Theoretical}\ }\textbf {\bibinfo {volume} {42}},\ \bibinfo
  {pages} {504007} (\bibinfo {year} {2009})}\BibitemShut {NoStop}%
\bibitem [{\citenamefont {Lemonik}\ and\ \citenamefont
  {Mitra}(2016)}]{Lemonik15}%
  \BibitemOpen
  \bibfield  {author} {\bibinfo {author} {\bibfnamefont {Y.}~\bibnamefont
  {Lemonik}}\ and\ \bibinfo {author} {\bibfnamefont {A.}~\bibnamefont
  {Mitra}},\ }\href {\doibase 10.1103/PhysRevB.94.024306} {\bibfield  {journal}
  {\bibinfo  {journal} {Phys. Rev. B}\ }\textbf {\bibinfo {volume} {94}},\
  \bibinfo {pages} {024306} (\bibinfo {year} {2016})}\BibitemShut {NoStop}%
\bibitem [{\citenamefont {Russomanno}\ and\ \citenamefont
  {Torre}(2016)}]{Torre16}%
  \BibitemOpen
  \bibfield  {author} {\bibinfo {author} {\bibfnamefont {A.}~\bibnamefont
  {Russomanno}}\ and\ \bibinfo {author} {\bibfnamefont {E.~G.~D.}\ \bibnamefont
  {Torre}},\ }\href {http://stacks.iop.org/0295-5075/115/i=3/a=30006}
  {\bibfield  {journal} {\bibinfo  {journal} {EPL (Europhysics Letters)}\
  }\textbf {\bibinfo {volume} {115}},\ \bibinfo {pages} {30006} (\bibinfo
  {year} {2016})}\BibitemShut {NoStop}%
\bibitem [{\citenamefont {Yates}\ \emph {et~al.}(2016)\citenamefont {Yates},
  \citenamefont {Lemonik},\ and\ \citenamefont {Mitra}}]{Yates16}%
  \BibitemOpen
  \bibfield  {author} {\bibinfo {author} {\bibfnamefont {D.~J.}\ \bibnamefont
  {Yates}}, \bibinfo {author} {\bibfnamefont {Y.}~\bibnamefont {Lemonik}}, \
  and\ \bibinfo {author} {\bibfnamefont {A.}~\bibnamefont {Mitra}},\ }\href
  {\doibase 10.1103/PhysRevB.94.205422} {\bibfield  {journal} {\bibinfo
  {journal} {Phys. Rev. B}\ }\textbf {\bibinfo {volume} {94}},\ \bibinfo
  {pages} {205422} (\bibinfo {year} {2016})}\BibitemShut {NoStop}%
\bibitem [{\citenamefont {Berdanier}\ \emph {et~al.}(2018)\citenamefont
  {Berdanier}, \citenamefont {Kolodrubetz}, \citenamefont {Parameswaran},\ and\
  \citenamefont {Vasseur}}]{Berdanier18}%
  \BibitemOpen
  \bibfield  {author} {\bibinfo {author} {\bibfnamefont {W.}~\bibnamefont
  {Berdanier}}, \bibinfo {author} {\bibfnamefont {M.}~\bibnamefont
  {Kolodrubetz}}, \bibinfo {author} {\bibfnamefont {S.}~\bibnamefont
  {Parameswaran}}, \ and\ \bibinfo {author} {\bibfnamefont {R.}~\bibnamefont
  {Vasseur}},\ }\href@noop {} {\bibfield  {journal} {\bibinfo  {journal}
  {arxiv:1803.00019}\ } (\bibinfo {year} {2018})}\BibitemShut {NoStop}%
\bibitem [{\citenamefont {Ryu}\ \emph {et~al.}(2010)\citenamefont {Ryu},
  \citenamefont {Schnyder}, \citenamefont {Furusaki},\ and\ \citenamefont
  {Ludwig}}]{Ryu10}%
  \BibitemOpen
  \bibfield  {author} {\bibinfo {author} {\bibfnamefont {S.}~\bibnamefont
  {Ryu}}, \bibinfo {author} {\bibfnamefont {A.~P.}\ \bibnamefont {Schnyder}},
  \bibinfo {author} {\bibfnamefont {A.}~\bibnamefont {Furusaki}}, \ and\
  \bibinfo {author} {\bibfnamefont {A.~W.~W.}\ \bibnamefont {Ludwig}},\ }\href
  {http://stacks.iop.org/1367-2630/12/i=6/a=065010} {\bibfield  {journal}
  {\bibinfo  {journal} {New Journal of Physics}\ }\textbf {\bibinfo {volume}
  {12}},\ \bibinfo {pages} {065010} (\bibinfo {year} {2010})}\BibitemShut
  {NoStop}%
\bibitem [{\citenamefont {Verresen}\ \emph {et~al.}(2017)\citenamefont
  {Verresen}, \citenamefont {Moessner},\ and\ \citenamefont
  {Pollmann}}]{Pollmann17}%
  \BibitemOpen
  \bibfield  {author} {\bibinfo {author} {\bibfnamefont {R.}~\bibnamefont
  {Verresen}}, \bibinfo {author} {\bibfnamefont {R.}~\bibnamefont {Moessner}},
  \ and\ \bibinfo {author} {\bibfnamefont {F.}~\bibnamefont {Pollmann}},\
  }\href {\doibase 10.1103/PhysRevB.96.165124} {\bibfield  {journal} {\bibinfo
  {journal} {Phys. Rev. B}\ }\textbf {\bibinfo {volume} {96}},\ \bibinfo
  {pages} {165124} (\bibinfo {year} {2017})}\BibitemShut {NoStop}%
\bibitem [{\citenamefont {Shirley}(1965)}]{Shirley65}%
  \BibitemOpen
  \bibfield  {author} {\bibinfo {author} {\bibfnamefont {J.~H.}\ \bibnamefont
  {Shirley}},\ }\href {\doibase 10.1103/PhysRev.138.B979} {\bibfield  {journal}
  {\bibinfo  {journal} {Phys. Rev.}\ }\textbf {\bibinfo {volume} {138}},\
  \bibinfo {pages} {B979} (\bibinfo {year} {1965})}\BibitemShut {NoStop}%
\bibitem [{\citenamefont {Sambe}(1973)}]{Sambe73}%
  \BibitemOpen
  \bibfield  {author} {\bibinfo {author} {\bibfnamefont {H.}~\bibnamefont
  {Sambe}},\ }\href {\doibase 10.1103/PhysRevA.7.2203} {\bibfield  {journal}
  {\bibinfo  {journal} {Phys. Rev. A}\ }\textbf {\bibinfo {volume} {7}},\
  \bibinfo {pages} {2203} (\bibinfo {year} {1973})}\BibitemShut {NoStop}%
\bibitem [{\citenamefont {Peschel}\ and\ \citenamefont
  {Eisler}(2009)}]{Eisler2009}%
  \BibitemOpen
  \bibfield  {author} {\bibinfo {author} {\bibfnamefont {I.}~\bibnamefont
  {Peschel}}\ and\ \bibinfo {author} {\bibfnamefont {V.}~\bibnamefont
  {Eisler}},\ }\href {http://stacks.iop.org/1751-8121/42/i=50/a=504003}
  {\bibfield  {journal} {\bibinfo  {journal} {Journal of Physics A:
  Mathematical and Theoretical}\ }\textbf {\bibinfo {volume} {42}},\ \bibinfo
  {pages} {504003} (\bibinfo {year} {2009})}\BibitemShut {NoStop}%
\bibitem [{\citenamefont {Amico}\ \emph {et~al.}(2008)\citenamefont {Amico},
  \citenamefont {Fazio}, \citenamefont {Osterloh},\ and\ \citenamefont
  {Vedral}}]{RMPVedral}%
  \BibitemOpen
  \bibfield  {author} {\bibinfo {author} {\bibfnamefont {L.}~\bibnamefont
  {Amico}}, \bibinfo {author} {\bibfnamefont {R.}~\bibnamefont {Fazio}},
  \bibinfo {author} {\bibfnamefont {A.}~\bibnamefont {Osterloh}}, \ and\
  \bibinfo {author} {\bibfnamefont {V.}~\bibnamefont {Vedral}},\ }\href
  {\doibase 10.1103/RevModPhys.80.517} {\bibfield  {journal} {\bibinfo
  {journal} {Rev. Mod. Phys.}\ }\textbf {\bibinfo {volume} {80}},\ \bibinfo
  {pages} {517} (\bibinfo {year} {2008})}\BibitemShut {NoStop}%
\bibitem [{Sup()}]{Suppmat}%
  \BibitemOpen
  \href@noop {} {}\bibinfo {note} {See Supplemental Material.}\BibitemShut
  {Stop}%
\bibitem [{\citenamefont {Wolf}(2006)}]{Wolf06}%
  \BibitemOpen
  \bibfield  {author} {\bibinfo {author} {\bibfnamefont {M.~M.}\ \bibnamefont
  {Wolf}},\ }\href {\doibase 10.1103/PhysRevLett.96.010404} {\bibfield
  {journal} {\bibinfo  {journal} {Phys. Rev. Lett.}\ }\textbf {\bibinfo
  {volume} {96}},\ \bibinfo {pages} {010404} (\bibinfo {year}
  {2006})}\BibitemShut {NoStop}%
\bibitem [{\citenamefont {Gioev}\ and\ \citenamefont {Klich}(2006)}]{Klich06}%
  \BibitemOpen
  \bibfield  {author} {\bibinfo {author} {\bibfnamefont {D.}~\bibnamefont
  {Gioev}}\ and\ \bibinfo {author} {\bibfnamefont {I.}~\bibnamefont {Klich}},\
  }\href {\doibase 10.1103/PhysRevLett.96.100503} {\bibfield  {journal}
  {\bibinfo  {journal} {Phys. Rev. Lett.}\ }\textbf {\bibinfo {volume} {96}},\
  \bibinfo {pages} {100503} (\bibinfo {year} {2006})}\BibitemShut {NoStop}%
\bibitem [{\citenamefont {Ares}\ \emph {et~al.}(2015)\citenamefont {Ares},
  \citenamefont {Esteve}, \citenamefont {Falceto},\ and\ \citenamefont
  {de~Queiroz}}]{ares15}%
  \BibitemOpen
  \bibfield  {author} {\bibinfo {author} {\bibfnamefont {F.}~\bibnamefont
  {Ares}}, \bibinfo {author} {\bibfnamefont {J.~G.}\ \bibnamefont {Esteve}},
  \bibinfo {author} {\bibfnamefont {F.}~\bibnamefont {Falceto}}, \ and\
  \bibinfo {author} {\bibfnamefont {A.~R.}\ \bibnamefont {de~Queiroz}},\ }\href
  {\doibase 10.1103/PhysRevA.92.042334} {\bibfield  {journal} {\bibinfo
  {journal} {Phys. Rev. A}\ }\textbf {\bibinfo {volume} {92}},\ \bibinfo
  {pages} {042334} (\bibinfo {year} {2015})}\BibitemShut {NoStop}%
\bibitem [{\citenamefont {Peschel}(2004)}]{Peschel04}%
  \BibitemOpen
  \bibfield  {author} {\bibinfo {author} {\bibfnamefont {I.}~\bibnamefont
  {Peschel}},\ }\href {http://stacks.iop.org/1742-5468/2004/i=06/a=P06004}
  {\bibfield  {journal} {\bibinfo  {journal} {Journal of Statistical Mechanics:
  Theory and Experiment}\ }\textbf {\bibinfo {volume} {2004}},\ \bibinfo
  {pages} {P06004} (\bibinfo {year} {2004})}\BibitemShut {NoStop}%
\end{thebibliography}
%

\begin{widetext}

\section{Supplementary Material}

This section contains:\newline
{\bf A: Explanation of why Majorana modes in ES are $|Z_0\pm Z_{\pi}|$ at $t=t^*$}\newline 
{\bf B: Explanation of why the strongest time-dependence is at zero entanglement energies}\newline
{\bf C: Convergence of central charge}\newline
{\bf D: Discontinuity in ${\cal M}_k$ for the static and Floquet system}\newline
{\bf E: Derivation of Entanglement Entropy of the Floquet Kitaev chain}

\subsection{A: Explanation of why Majorana modes in ES are $|Z_0\pm Z_{\pi}|$ at $t=t^*$}

To understand why the number of Majorana edge modes is \(|Z_0 \pm
Z_{\pi}|\) at the two time-reversal-symmetric (TRS) points $t^*$, it is helpful to
revisit how the two topological indices \(Z_0, Z_\pi\) are
calculated \cite{Delplace14,Asboth13}. 

The static Hamiltonian 
belongs to class BDI. Denoting \(\mathcal{K}\) as complex conjugation, time-reversal
symmetry corresponds to \(\mathcal{T} = \mathcal{K}\), particle-hole symmetry to
\(\mathcal{P} = \left(  \sigma_x \otimes 1_N \right) \mathcal{K}\), and chiral symmetry to \(
\Gamma = \mathcal{T}\cdot \mathcal{P} = \sigma_x \otimes 1_N \). Note that $\Gamma^2=1$,
and $N$ denotes the number of sites.

For the Floquet system, PHS symmetry is obeyed at every instant of time.
However chiral symmetry and TRS for a Floquet system is more subtle since \(H(t) \ne H(-t)\). In particular TRS holds only at 
the two times $t^*=T/4, 3T/4$. We now discuss the windings at these two times. 

The evolution operator over one drive cycle will produce an effective
Hamiltonian, \( H_{\rm eff}(t^*)\), which depends on the starting time of the period,
\begin{equation*}
  U(t^*) = \mathbb{T} e^{-i \int_{t^*}^{t^* + T} dt' H(t')} = e^{-i H_{\rm eff}(t^*)T}.
\end{equation*}
The effective Hamiltonian, \( H_{\rm eff}(t^*)\), describes stroboscopic
evolution, and its spectrum yields the quasienergies,  
\( U(t^*)^n | \phi_\epsilon \rangle = e^{-i \epsilon T n} |\phi_\epsilon
\rangle \).

Chiral symmetry of a periodically driven system is the statement that there
exists a starting time \(t^*\) such that,
\begin{equation*}
  \Gamma H_{\rm eff}(t^*) \Gamma = -H_{\rm eff}(t^*) \leftrightarrow \Gamma U(t^*) \Gamma = U(t^*)^{-1}. 
\end{equation*}
This definition allows for \( H_{\rm eff}(t^*)\) to have a well defined winding.

Chiral symmetry can also be phrased as the ability to find an intermediate time
\(t_0\), such that,
\begin{align*}
  &U'=U(t^*) = F_2 F_1 = \Gamma F^\dagger \Gamma F & \text{where }F_1= F= \mathbb{T}e^{-i \int_{t^*}^{t^*+ t_0} dt'H(t')},\ 
                                                   F_2 = \Gamma F^\dagger \Gamma =\mathbb{T}e^{-i \int_{t^*+t_0}^{t^* + T} dt' H(t')}
  \end{align*} 
  We can also find the time-shifted propagator, \( U'' = F_1 F_2 = F \Gamma
  F^\dagger \Gamma \). It is easy to check both of these propagators satisfy the
  above definition of chiral symmetry. The shifted propagator $U''$ corresponds to
  picking the other TRS point as the starting time for our period. We will work
  in the basis where \( \Gamma \) is diagonal. This is equivalent to the
  complex fermion to Majorana basis transformation. 

 In the diagonal basis, \(\Gamma\) takes the
  form \( \sigma_z\otimes 1_N\), where $\otimes 1_N$ denotes that \(\sigma_z\) acts on each physical
  site. \( \Gamma\) acting on A sites will yield 1, while acting on
  B sites will yield -1. Thus a wavefunction that resides on both
  sublattices will not be invariant under the action of \(\Gamma\), while states
  that reside only on one of the two sublattices will only pick up a phase. 

  Suppose \( | \Psi'_{\epsilon}\rangle \) is an eigenstate of $H_{\rm eff}(t^*)$
  with eigenvalue $\epsilon$. This state has a chiral symmetric partner \(
  \Gamma |\Psi'\rangle \) with eigenvalue \( -\epsilon \). At \( \epsilon = 0,
  \pi \), we can form the states \( \frac{1}{\sqrt{2}}\left( |\Psi_{0/\pi}'
    \rangle \pm \Gamma |\Psi_{0/\pi}'\rangle \right)\), which are eigenstates of
  \( \Gamma \), and thus reside on one of the sublattices.

  So with \( U', U''\), we have two effective Hamiltonians each with well
  defined windings, \( \nu' , \nu'' \). We can break down these windings into
  the number of edge modes living on one of the edges, with support only on sublattice A/B, and
  with quasienergy 0/\(\pi\), as follows,
  \begin{align*}
    \nu' &= n_{A,0}' -n_{B,0}' + n_{A,\pi}' - n_{B,\pi}' \\
    \nu''&= n_{A,0}''-n_{B,0}''+ n_{A,\pi}''- n_{B,\pi}''. 
  \end{align*}
One way to see the \( \nu = n_A - n_B\) breakdown is as follows. The
winding \(\nu\) is defined via an integral over the BZ. One can reverse the
sign of \(\nu\) by a spatial inversion which is equivalent to interchanging the 
A and B sublattice.

Finally, consider the state \( |\Psi_\epsilon'\rangle \), where \(U' |\Psi_\epsilon' \rangle =
 e^{-i \epsilon}|\Psi_\epsilon'\rangle \), with \( \epsilon = 0,\pi\). As we
 already discussed, it resides on one of the two sublattices, \( \Gamma
 |\Psi_{\epsilon}'\rangle = e^{-i \gamma}|\Psi_\epsilon'\rangle \), with
 \(\gamma = 0, \pi\), for the A/B sublattices. Now consider the state \( |\Psi_\epsilon''
 \rangle = F |\Psi_\epsilon'\rangle \). This is an eigenstate of \( U'' \), with the
 same quasienergy. This state is also on a single sublattice,
 \begin{equation*}
   \Gamma |\Psi_\epsilon'' \rangle = \Gamma F \Gamma e^{i \gamma} |\Psi_\epsilon' \rangle
   = \Gamma F \Gamma e^{i( \gamma +\epsilon)} U' |\Psi_\epsilon'\rangle =
   e^{i (\gamma +\epsilon)} |\Psi_\epsilon''\rangle.
   \end{equation*}
The above shows that the phase picked up on application of $\Gamma$ depends on the quasienergy of the state. 
\( | \Psi_\epsilon''\rangle \) is on the same sublattice as \(
|\Psi_\epsilon'\rangle \) if \(
\epsilon = 0\), and on the opposite sublattice if \( \epsilon = \pi \). This is to
say, \( n_{A,\pi}'' = n_{B,\pi}', n_{A,\pi}' = n_{B,\pi}'', n_{A,0}'' =
n_{A,0}', n_{B,0}'' = n_{B,0}'\), which when plugged into the above relation for
\( \nu', \nu'' \) gives us, 
\begin{align*}
  Z_0 &= \frac{n_{A,0}'  - n_{B,0}'+ n_{A,0}'' -n_{B,0}''}{2} = n_{A,0}'  - n_{B,0}'= \frac{1}{2} \left(  \nu' + \nu'' \right)\\
  Z_\pi&=  \frac{n_{A,\pi}' - n_{B,\pi}' + n_{B,\pi}''-n_{A,\pi}''}{2} = n_{A,\pi}' - n_{B,\pi}' =\frac{1}{2} \left( \nu' - \nu'' \right).
\end{align*}
From above we see that the number of Majorana modes at one TRS time is $\nu'= Z_0+Z_{\pi}$, and at the other
it is $\nu'' = Z_0-Z_{\pi}$. Since the entanglement spectrum at TRS times is constructed from $|\Psi'\rangle, |\Psi''\rangle$,
it inherits the same properties.

\subsection{B: Explanation of why the strongest time-dependence is at zero entanglement energies}

Note that the Schmidt states with zero entanglement energies correspond to
the Majorana modes that are localized at the entanglement cut. This can be noted from
the fact that zero entanglement energies contribute more to the entanglement entropy, and therefore
must arise from boundary states. 
 
Floquet micromotion
causes these modes to hybridze with each other during
times away from TRS times i.e, $(t\neq t^*)$, while they are uncoupled only at $t=t^*$. 
All the time-dependence comes from these modes hybridizing
and unhybridizing where the hybridization shifts their entanglement energies symmetrically
around zero.  

The states with large (in magnitude) entanglement energies are bulk states which are 
almost uniformly distributed within the cut. Their micromotion causes relatively
smaller fractional changes to their entanglement energies, and therefore appear to be almost
time-independent.

\subsection{C: Convergence of the central charge}

\begin{figure}[h!]
    \centering
\includegraphics[width=0.5\textwidth,keepaspectratio]{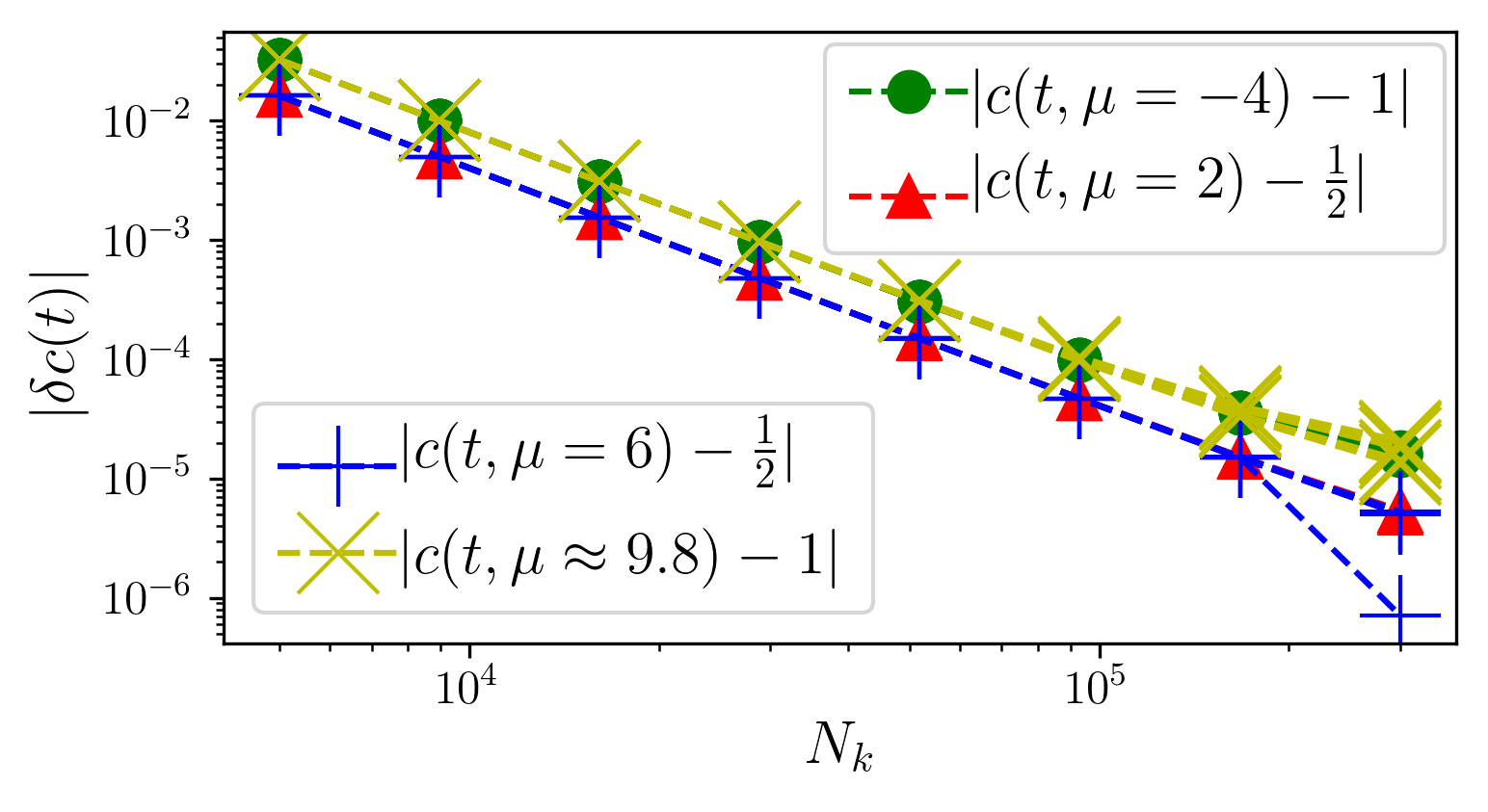}
\caption{
    Deviation of central charge from Eq. (5) for five different times within a
period, and for four different critical $\mu$, 
plotted against the resolution of the Brillouin zone. The lines for different times are indistinguishable.
    The fit function is $f(L) = a + \frac{b}{L} + \frac{c}{3} \log (L)$, where $400 \leq L\leq 600$.
}\label{central}
\end{figure}

\subsection{D: Discontinuity in ${\cal M}_k$ for the static and Floquet system}

Here we discuss the ``Fermi''-points of the static BdG system and we give
additional examples for the Floquet analog already discussed in the main text.
Fig.~\ref{staticjumps} shows the discontinuities at two values of $\mu$ that are
tuned at the critical point of the static chain. The jump is always across the
Bloch sphere of radius equal to the discontinuity in the eigenvalues of ${\cal
  M}_k$, which is $2$. Due to TRS, $\mathcal{M}_k$ cannot have a projection onto
$\sigma_x$.

    \begin{figure}[h!]
        \centering
\includegraphics[width=0.5\textwidth,keepaspectratio]{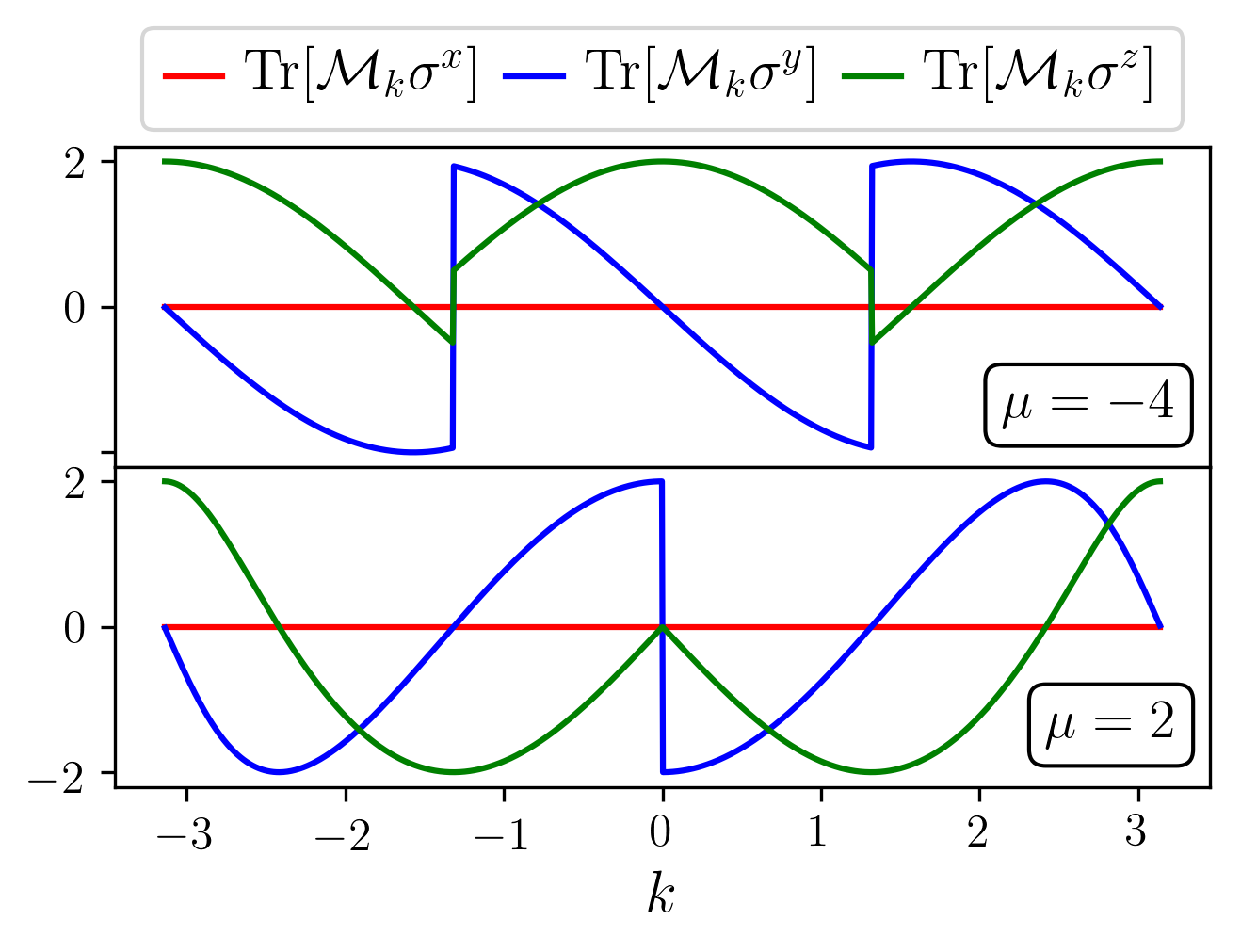}
\caption{Discontinuities in the $\mathcal{M}_{k,\rm static}$ matrix for the
  static system, at the critical $\mu$ points. The discontinuities send
  $\vec{d}$ to its opposite location on the Bloch-sphere of radius 2. All the
  projections are continuous for $\mu$ away from the critical values. Top panel
  with $\mu=-4$ has $N_T=2\times2=4$ and a central charge of $c=N_T/4=1$. The
  lower panel with $\mu=2$ has $N_T=2$, and a central charge $c=1/2$.
}\label{staticjumps}
\end{figure}

In the Floquet setting, the discontinuities in \(\mathcal{M}_{k,\rm FGS}\) now
have a time dependent orientation, but the strength of the discontinuities is
fixed at 2, jumping across the diameter of the Bloch-sphere.
Fig.~\ref{floquetjumps} shows two more examples of phase transitions, a \((1,0)
\rightarrow (0,0)\) transition and a \( (0,0)\rightarrow (0,-2)\). 
\begin{figure}[h!]
  \centering
\includegraphics[width = 0.6\textwidth,keepaspectratio]{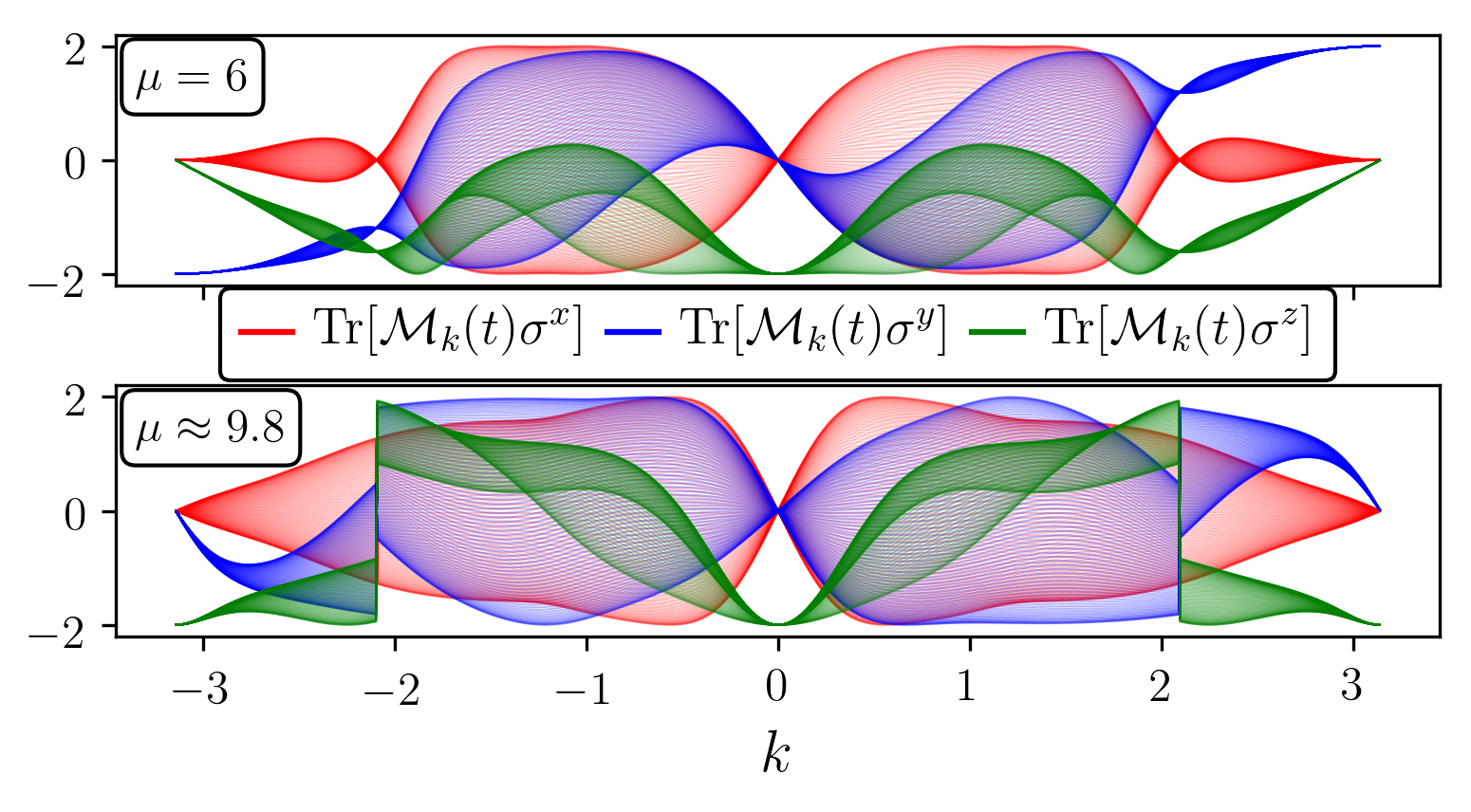}
\caption{Discontinuities in \(\mathcal{M}_{k,\rm FGS}(t)\) for many times during
  a driving period (different solid lines) at two different critical \(\mu\).
  The discontinuities send the Bloch-vector to the opposite side of the sphere
  at all times. The top panel is showing the discontinuities for the \( (1,0)
  \rightarrow (0,0)\) transition, which has a gap closing at \( k = \pi\).
  For this special case, the orientation of the jump remains fixed in time because the
  Hamiltonian becomes time independent at \( k = \pi\). The bottom panel
  highlights the case of a MPM transition, \( (0,0) \rightarrow (0,-2)\). Here,
  the Bloch-vector is flipped across the origin, and has components in all three
  directions. The discontinuities in  ${\rm Tr}\left[{\cal M}_{k,{\rm FGS}}(t)\sigma_x\right]$ are difficult
  to see. 
}
\label{floquetjumps}
\end{figure}

\subsection{E: Derivation of Entanglement Entropy of the Floquet Kitaev chain}
Using Eq.~(6) in the main text, we write the correlation matrix (keeping the singular part of ${\cal M}_{k,\rm FGS}$) as,
\begin{align}\label{corrS}
&G_{i,j}(t)=\int_{-\pi}^{\pi}\frac{dk}{2\pi}e^{i k (i-j)}\biggl[\frac{k-k^*}{|k-k^*|} {\sigma}_{1}(t)\biggr]
\sim e^{ik^*(i-j)}\frac{i}{\pi (i-j)}{\sigma}_{1}(t)
\end{align}
Then we need to find the eigenspectrum of,
\begin{equation}
\sum_jG_{i,j}\begin{pmatrix}\phi_j \\ \psi_j\end{pmatrix} = \lambda \begin{pmatrix}\phi_i \\ \psi_i\end{pmatrix}
\end{equation}
We now define,
\begin{eqnarray}
\Psi_j = e^{-ik^*j}\psi_j; \,\,\, \Phi_j = e^{-ik^*j} \phi_j
\end{eqnarray}
We find it convenient to perform a unitary rotation at every time $t$ so as to align $\vec{\sigma}_1(t)$ along $\sigma_x$. This
does not affect the eigenvalues. The eigenvectors have a time-dependence, and we do not show it explicitly.
Then,
\begin{eqnarray}
&&i\frac{1}{\pi}\sum_{j=1\ldots N} \frac{1}{i-j}\Phi_j  = \lambda \Psi_i \Rightarrow -iD_{ij}\Phi_j = \lambda \Psi_i \label{eqP}\\
&&i\frac{1}{\pi}\sum_{j=1\ldots N}\frac{1}{i-j}\Psi_j  = \lambda \Phi_i \Rightarrow -i{D}_{ij}\Psi_j = \lambda \Phi_i\label{eqS}
\end{eqnarray}
where,
\begin{eqnarray}
&&D_{ij}= \frac{1}{\pi}\biggl(\frac{1}{i-j}\biggr)\label{Ddef}
\end{eqnarray}
Thus we have a combined equation
\begin{eqnarray}
{D}_{il}D_{lj}\Phi_j = -\lambda^2 \Phi_i
\end{eqnarray}
This may be solved following for example the approach in Ref.~\onlinecite{Peschel04}. We give the details for
completeness.

The above may be recast as
\begin{eqnarray}
4\sum_{l\neq i=1\dots N}K_{il}\Phi_l =\biggl(\lambda^2-4 K_{ii}\biggr)\Phi_i
\end{eqnarray}
with $K_{il}=-{D}_{ij}D_{jl}$ defined as
\begin{eqnarray}
{K}_{il}=  -\sum_{j} \frac{1}{\pi\left(2(i-j)\right)}\frac{1}{\pi\left(2(j-l)\right)}=-\frac{1}{2\pi^2(i-l)}
\sum_j\biggl[\frac{1}{2(i-j)} +\frac{1}{2(j-l)}\biggr]={K}_{li}\label{K1adef}\\
{K}_{ii} = \sum_{j=1\ldots N}\frac{1}{\pi^2\left(2(i-j)-1\right)^2}\label{K2adef}
\end{eqnarray}
where for $K_{ii}$ we place a short-distance cutoff to regularize the integral.

As long as $i$ is not too close to the boundaries,
and $N$ is large enough, $K_{ii}=\sum_{j=1\ldots N}\frac{1}{\pi^2 (2j-2i+1)^2} =1/4$.
The non-local term can be approximated by an integral, where $Nx_{l}= 2l-1 $ and the volume of one point is $\Delta x = 2/N$. Thus,
\begin{eqnarray}
&&K_{il} = -\frac{1}{2\pi^2(i-l)}
\sum_{j=1}^{N}\biggl[\frac{1}{2(i-j)} +\frac{1}{2(j-l)}\biggr]\nonumber\\
&&= -\frac{1/N}{\pi^2(x_i-x_l)}
\sum_{x_j=1/N\ldots (2-1/N)}\biggl[\frac{1}{N(x_i-x_j)} +\frac{1}{N(x_j-x_l)}\biggr]\nonumber\\
&&\sim -\frac{1/N}{\pi^2(x_i-x_l)}\int_{1/N}^{2-1/N} \frac{dx_j}{\Delta x}\biggl[\frac{1}{N(x_i-x_j)} +\frac{1}{N(x_j-x_l)}\biggr]
\end{eqnarray}

Performing the integral, we obtain, in the large $N$ limit,
\begin{eqnarray}
K_{il} =-\frac{1/N}{2\pi^2(x_i-x_l)} \biggl[\ln\biggl(\frac{x_l-2}{x_l}\biggr)
-\ln\biggl(\frac{x_i-2}{x_i}\biggr)  \biggr]
\end{eqnarray}
Thus we have to solve the eigenvalue equation, (where we use that $\sum_l = \frac{N}{2}\int dx'$)
\begin{eqnarray}
&&\frac{1}{2}\int_{1/N}^{2-1/N} dx'K(x,x')\Phi(x')= \biggl(\frac{\lambda^2}{4}- K_{ii}\biggr)\Phi(x)\\
&&K(x,x') = -\frac{1}{2\pi^2(x-x')} \biggl[\ln\biggl(\frac{x}{2-x}\biggr)
-\ln\biggl(\frac{x'}{2-x'}\biggr)  \biggr]
\end{eqnarray}
and,
\begin{eqnarray}
\Psi(x) = \pm \Phi(x)
\end{eqnarray}

Changing variables to
\begin{eqnarray}
&&u(x) = \frac{1}{2}\ln\biggl(\frac{x}{2-x}\biggr) \Rightarrow \,\, \frac{du}{dx} = \frac{1}{t(x)}; t(x) = x(2-x)\\
&& x =2\frac{e^{u}}{e^{u}+e^{-u}};\,\, \sqrt{t(x)} = \frac{1}{\cosh(u)}\\
&& x-x' = \frac{\sinh(u-u')}{\cosh(u)\cosh(u')}
\end{eqnarray}
Thus, defining
\begin{eqnarray}
\Phi(u) = \frac{\chi(u)}{\sqrt{t(u)}}= \chi(u)\cosh(u)
\end{eqnarray}
we need to solve the eigenvalue problem,
\begin{eqnarray}
&&-\frac{1}{2\pi^2}\int_{-\infty}^{\infty} du'\frac{u-u'}{\sinh(u-u')} \chi(u') =\biggl(\frac{\lambda^2}{4}- K_{ii}\biggr) \chi(u)\\
\end{eqnarray}
Using that,
\begin{eqnarray}
\int_{-\infty}^{\infty} dx \frac{x}{\sinh(x)}e^{-i q x} = \frac{\pi^2}{1+\cosh(\pi q)}=\frac{\pi^2}{2\cosh^2(\pi q/2)}
\end{eqnarray}

Thus,
\begin{eqnarray}
\lambda_q = \pm \tanh\left(\pi q/2\right) \label{ente}
\end{eqnarray}
Now we need to apply the boundary conditions.
Since the states have to be eigenstates of parity, we have
\begin{eqnarray}
\Phi_1 = \pm\Phi_N.
\end{eqnarray}
Since $t(x_1)=t(x_N)$, and $\chi =e^{iqu}$, we have,
\begin{eqnarray}
\frac{q}{2}\ln\biggl(\frac{x_{N}}{2-x_{N}}\biggr)- \frac{q}{2}\ln\biggl(\frac{x_{1}}{2-x_{1}}\biggr) = n\pi \label{bc2a}
\end{eqnarray}

This gives,
\begin{eqnarray}
q \ln{N} = n \pi \Rightarrow q = \frac{n\pi}{\ln{N}}
\end{eqnarray}
The entanglement energy is
\begin{eqnarray} \label{ent2}
&&S = -\frac{1}{2}\sum_{\lambda'}\biggl[\lambda'\ln(\lambda')+ (1-\lambda')\ln(1-\lambda')\biggr];\,\,
\epsilon =\ln\biggl[\frac{1-\lambda'}{\lambda'}\biggr]; \lambda' =\frac{1-\lambda}{2}.
\end{eqnarray}
Since from Eq.~\eqref{ente},
\begin{eqnarray}
\epsilon = \pi q = \frac{n\pi^2}{\ln{N}}=\alpha n; \,\, \alpha = \pi^2/\ln{N}
\end{eqnarray}
in terms of $\epsilon$,
\begin{eqnarray}
&&S= \frac{1}{2}\sum_{\epsilon }\biggl[\frac{\epsilon}{1+e^{\epsilon}} +\log\left(1+e^{-\epsilon}\right)\biggr]
\end{eqnarray}
Converting the sum into an integral, noting that,
\begin{eqnarray}
&&\int_{-\infty}^{\infty} dx \biggl[\frac{x}{1+e^{x}}+ \log\left(1+e^{-x}\right)\biggr]= \frac{\pi^2}{3}
\end{eqnarray}
we obtain the following result for the entanglement entropy,
\begin{eqnarray}
S=\frac{\pi^2}{6\alpha} = \frac{1}{6}\ln{N}
\end{eqnarray}
Thus the central charge is $c=1/2$ for the critical Floquet state with a single singular point (at $k=k^*$).

\end{widetext}

\end{document}